\documentclass[prb,twocolumn,nofootinbib]{revtex4}
\usepackage{natbib}
\usepackage{graphicx}
\usepackage{amssymb}

\usepackage{amsmath,amssymb,amsfonts,times,graphicx}
\usepackage[ps2pdf,bookmarks=true,colorlinks,linkcolor=red,urlcolor=blue,citecolor=blue]{hyperref}
\usepackage[usenames]{color}
\usepackage{dcolumn}

\begin{document}
\title{Experimental signatures of 3d fractional topological insulators}
\author{Brian Swingle}
\email{brians@physics.harvard.edu}
\affiliation{Department of Physics, Harvard University, Cambridge, MA 02138}
\begin{abstract}
In this work we explore experimental signatures of fractional topological insulators in three dimensions.  These are states of matter with a fully gapped bulk that host exotic gapless surface states and fractionally charged quasiparticles.  They are partially characterized by a non-trivial magneto-electric response while preserving time reversal.  We describe how these phases appear in a variety of probes including photoemmission, tunneling, and quantum oscillations.  We also discuss the effects of doping and proximate superconductivity.  We argue that despite our current theoretical inability to predict materials where such phases will realized, they should be relatively easy to detect experimentally.
\end{abstract}
\maketitle

\section{Introduction} 
Since the discovery of the fractional quantum Hall fluids \cite{fqhe,laughlin} and the subsequent elucidation of their topological structure \cite{toporder2,toporder1}, it has become clear that entanglement plays a crucial role in wide variety of zero temperature quantum phases of matter.  To describe such phases of matter, it is necessary to understand their pattern of long range entanglement since they fail to be distinguished by any symmetry breaking pattern.  This has led to a fruitful characterization of these phases in terms of a ground state wavefunction which cannot be adiabatically deformed into an unentangled state \cite{topent1,topent2}.  More recently, there has been an explosion of interest in topological insulators \cite{ti1,ti2,ti3,ti4,ti5,ti6,ti7,ti8,ti9,ti10}, band insulators where, perhaps depending on a symmetry, the ground state wavefunction is also not adiabatically connected to a product state (see Refs. \onlinecite{ti_rmp_2010_qi,ti_rmp_2010_kane} for reviews).  Unlike fractional quantum Hall fluids, interactions are not crucial to stabilize topological insulators. Instead, interactions lead to new exotic states in 3d which have highly anomalous surface states \cite{fti1,fti2}.  In this paper we describe a large number of robust experimental signatures of the simplest class of fractional topological insulators.

Let us briefly review the experimental situation with respect to fractionalized phases in more than one dimension.  The best examples remain the plethora of exotic states realizing fractional quantum Hall physics in 2d.  Fractional quantum Hall states are always gapped in the bulk (modulo certain interesting exceptions e.g. $\nu = 1/2$ \cite{hlr}) and support emergent fractionally charged anyons \cite{tqc_rmp}.  More recently, experimental evidence has accumulated for a different kind of fractionalized state in certain layered organic salts \cite{Yamashita04062010,PhysRevB.77.104413}.  In these states the charge motion is frozen but the spins continue to fluctuate and appear to form a liquid-like state down to low temperatures.  This is all extremely exciting.  Importantly, both these classes of examples have a common element: they possess gapless modes either in the bulk or at the edge.  In 3d we currently have no well established experimental examples of fractionalized phases.  However, we have stable field theories\cite{fti1,fti2,fti_field} and exactly solvable models\cite{fti_exact} that demonstrate that phases of the type we consider below can exist.  On the other hand, many other fractionalized phases are possible that have no gapless modes in their spectrum even at a boundary\cite{stringnet}. Enter fractional topological insulators.  Precisely because they have fractionalized time reversal protected surface states they are much more experimentally accessible compared to their gapped cousins.  We think it likely that the first experimental examples of fractionalized phases in 3d will either be gapless in the bulk or have gapless surface states, so we believe it is of considerable interest to explore the basic experimental properties of fractional topological insulators.

We study time reversal invariant phases of electrons in three dimensions that realize fractional topological insulators \cite{fti1,fti2}.  Such a state is an electric and thermal insulator with a gap to all electronic excitations and collective modes.  As long as time reversal is not broken, the state is stable to all zero temperature perturbations (in fact, the bulk state remains non-trivial even without $\mathcal{T}$, but the surface states may disappear).  It contains a $\theta$ term $\theta \vec{E}\cdot \vec{B}$ in its low energy effective action that describes a fractional magneto-electric effect relating, among other things, polarization to magnetization.  If the sample is terminated at a surface, then that surface carries protected gapless states described by weakly interacting quasiparticles of \textit{fractional} electric charge.  Despite the fact that the bulk is insulating, the presence of surface states can still lead to electrical conduction.  This can occur either as a pure surface effect or, for example, because polycrystalline samples might have many low lying states in the putative bulk that can participate in conduction.

There are by now many experimental examples of weakly correlated topological insulators in the non-trivial $Z_2$ class in three dimensions \cite{ti_arpes_2008,tunable_ti_2009,largegap_ti_2009,doped_ti_2010}.  A wide variety of experimental tools have been applied to these systems, including photoemission \cite{ti_arpes_2008,largegap_ti_2009}, STM \cite{stm_ti1,stm_ti2}, and quantum oscillations \cite{quantum_osc_bi2se3,quantum_osc_bi2te3,cyclotron_ti}.  We can also consider direct probes of the magneto-electric effect, for example, a giant Kerr rotation has been observed in Bi$_2$Se$_3$ \cite{giantkerr_ti}.  In addition, doping in the bulk and on the surface has proven to be useful control parameter as well as being present naturally in many experimental realizations \cite{tunable_ti_2009,quantum_osc_bi2se3}.  There have also been interesting considerations involving the superconducting proximity effect at the surface of a topological insulator \cite{sc_ti_interface,doped_ti_2010}.  In this paper, we focus on probes like photoemission, tunneling, and quantum oscillations since they give qualitatively distinct signatures and have already been brought to bear on weakly correlated topological insulators.  Of course, the magneto-electric effect is one of the defining characteristics of a fractional topological insulator, but fractional topological insulators differ primarily quantitatively from ordinary topological insulators in these probes.  Thus we choose to focus on probes where there is a qualitative distinction between the two phases.

There are a number of other reasons to study experimental signatures of fractional topological insulators even though we lack an experimental example.  First, while there have now been impressive theoretical predictions of materials entering a topological insulator phase, the analogous problem for fractional topological insulators is quite difficult.  Since interactions play a crucial role, the many advanced tools available to analyze and predict effective band structures for weakly correlated phases are not directly applicable.  Mean field theory also totally fails to capture fractional topological insulators since they lack a local order parameter.  Exact diagonalization and DMRG, which are so powerful in one and even two dimensions, are nearly hopeless in three dimensions.  Various quantum Monte Carlo approaches may be useful, but the models which realize fractional topological insulators are likely to have a sign problem.  We can produce trial wavefunctions for these phases \cite{fti1,fti2}, so variational Monte Carlo could be useful, but it hardly paints a conclusive picture.  Thus we believe that when such materials are discovered, it will to some extent be experimental chance, and hence we should be as clear as possible about the various experimental signatures.  Second, there are other gapped states of electrons, also fractionalized, which do not possess protected surface states.  These states are even harder to detect experimentally, but as we will describe, some may be connected via a second order phase transition to a fractional topological insulator.  Hence an experimental identification of a fractional topological insulator would also potentially provide a root to a proximate fractional phase without surface states for which there are also no known experimental examples.

This paper is organized as follows.  First we review and refine the general theory introduced in Refs. \onlinecite{fti1,fti2} including the effects of superconductivity and doping.  Since we do not know the detailed experimental situation, we try to be general and touch on a wide variety of possibilities. Then we describe in turn signatures in photoemission, STM, and quantum oscillations.  Finally, we give a concluding discussion.  Readers interested in a brief summary of the results may skip to the discussion.

\section{Basic theoretical description} For concreteness we focus on the simplest electronic fractional topological insulator.  Quite analagous to the $\nu = 1/3$ Laughlin state \cite{laughlin}, the electrons in this fractional topological insulator have fractionalized into charge $e/3$ constituents.  We may imagine a $1/3$ partially filled topological band subject to strong electron interactions.  Naturally, many conventional metallic and symmetry breaking states are possible in such a situation, but more exotic states, analogous to fractional quantum Hall liquids are also possible. Conventional states may also coexist with exotic structures.  To access these states we write
\begin{equation}
c_r = f_{r 1} f_{r 2} f_{r 3}
\label{su3-parton}
\end{equation}
where $c$ is an electron operator and the $f$s are fractionally charged ``partons"\cite{wen_partons}.  Parton models of fractionalized phases are well known in the fractional quantum Hall effect and can capture the universal features of the low energy physics.  The parton decomposition is illustrated in Fig. 1.  Note that we consider phases that incorporate strong spin-orbit coupling, hence we suppress whenever possible the explicit spin label on the electrons.  If we treat the $f$s as fermions, then it is a simple algebraic exercise to verify that $c$ and $c^+$ also satisfy fermionic commutation relations provided certain constraints are satisfied.  These constraints, which reduce the unphysical Hilbert space of generated by $f_a$ to the physical Hilbert space generated by $c$, are implemented by coupling the $f_a$ to an emergent gauge field. 

The ansatz in Eq. \ref{su3-parton} has an $SU(3)$ redundancy, but we do not deal directly with such non-Abelian gauge fields; instead we consider a phase where the $SU(3)$ is broken to its center $Z_3$.  Practically speaking, this breaking of $SU(3) \rightarrow Z_3$ mixes the parton colors so that we no longer need label partons by their color.  We should also emphasize that even if the system enters a fractionalized phase where the partons are deconfined, the physical parton excitations will inevitably be coarse grained and renormalized versions of the lattice scale partons we formally introduced here.

If we insist on keeping track of the spin degree of freedom, as will be important when considering magnetic perturbations, the following parton construction, a refinement of Eq. \ref{su3-parton} is more appropriate.  At every site we write
\begin{eqnarray}
c_\uparrow = f_\uparrow d_\uparrow d_\downarrow \nonumber \\
c_\downarrow = f_\downarrow d_\uparrow d_\downarrow
\label{spin-parton}
\end{eqnarray}
in terms of two species of partons $f$ and $d$.  The decomposition in Eq. \ref{spin-parton} has a $U(1)$ redundancy in which the $d$ fermions have charge $1$ while the $f$ fermions have charge $-2$.  This insures that $c$ is gauge invariant.  $d$ and $f$ are taken to have equal electromagnetic charge $e/3$.  Bosonic terms of the form $f^\dagger d$ in the low energy Hamiltonian mix the $d$ and $f$ fermions and break the $U(1)$ gauge redundancy to a $Z_3$ subgroup (the operator $f^\dagger d$ carries charge $3$).  There will also be spin non-conserving terms in the low energy Hamiltonian.  Note that if the $f_\sigma$ and $d_\sigma$ fermions transform in the usual way under time reversal i.e. $f_\uparrow \rightarrow f_\downarrow$ and $f_\downarrow \rightarrow - f_\uparrow$ then the physical electron operators also transform in the same way.  Because the $d$s and $f$s are mixed, we henceforth refer to all partons using the symbol $f$.  These fermions carry electromagnetic charge $e/3$, $Z_3$ charge 1, and spin $1/2$ (which is broken by their bandstructure).  Now we give the basic physical description of a fractional topological insulator.

\begin{figure}
\begin{center}
\includegraphics[width=.48\textwidth]{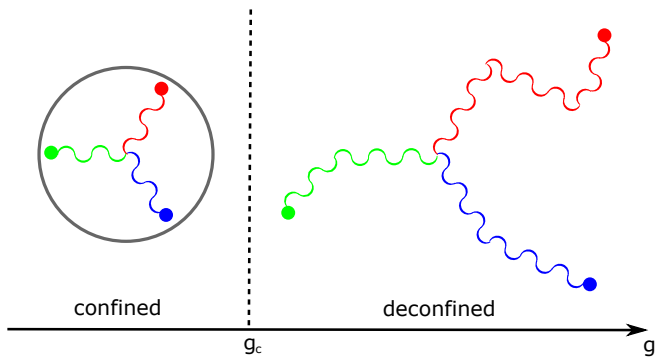}
\end{center}
\caption{The three colored balls each correspond to a different parton type $f_a$ (in the $SU(3)$ language).  In a weakly correlated phase, shown on the left, the partons are confined by the gauge field (wiggly lines) into an electron (gray circle).  This confinement is typically so tight that there is little meaning to the individual partons.  However, a deconfined phase (shown on the right) is also possible where the partons are relatively free to roam, although still connected through an emergent gauge field.  Typically we can realize both types of phases as a function of some control parameter $g$.  The confinement/deconfinement quantum phase transition at $g_c$ is often first order, but it may also be second order.}
\end{figure}

A convenient starting point to describe the physics of this state is a $Z_3$ lattice gauge theory describing a renormalized system with generic interactions (see Ref. \onlinecite{strongcoupling} for an introduction in the context of lattice quantum Hall states).  We could also work directly in the continuum, but the lattice setting is more natural from the point of view of discrete gauge theory.  The ingredients are the partons $f_{r \sigma}$ ($\sigma$ is a spin label), the gauge fields $z_{rr'}=1,e^{2 \pi i/3}, e^{4 \pi i/3}$, and the dual electric fields $e_{rr'} = 0,1,2$.  The gauge fields are oriented and we take $z_{r'r} = z_{rr'}^*$. The partons carry one unit of $Z_3$ charge and the fundamental gauge field commutator is
\begin{equation}
e^{2 \pi i e_{rr'}/3} z_{rr'}= e^{2 \pi i/3} z_{rr'} e^{2 \pi i e_{rr'}/3}
\end{equation}
which says that the operator $z_{rr'}$ adds one unit of electric flux $e_{rr'}$.  The Hamiltonian is
\begin{eqnarray}\label{partonH}
H = - \sum_{rr'\sigma \sigma'} w_{rr'\sigma \sigma'} f^+_{r\sigma} z_{rr'} f_{r' \sigma'} \nonumber \\ + h \sum_{<rr'>} \sin^2{\left(2 \pi e_{rr'}/3\right)} - K \sum_{\text{loops}} \prod_{<rr'> \in \text{loop}} z_{rr'}
\end{eqnarray}
and is defined in terms of links $<rr'>$ and loops on some unspecified underlying graph.  We impose the Gauss' law constraint that electric flux lines only end on charges.  

We are interested in the deconfined phase of Eq. \ref{partonH} that is obtained when $K,w \gg h$.  The band structure $w_{rr'\sigma \sigma'}$ and filling of the $f$s is chosen so that they form a $Z_2$ non-trivial topological insulator in the limit of no $Z_3$ gauge fluctuations.  We can also include direct interactions between the partons, but these don't strongly modify the story (the gauge fields already generate effective interactions).  Now what is the physics of the deconfined phase of the Hamiltonian Eq. \ref{partonH}?

The bulk is fully gapped with excitations consisting of gapped charge $e^* = e/3$ fermionic quasiparticles, the $f$s, coupled to an emergent $Z_3$ gauge field.  Such a discrete gauge field may be understood heuristically as arising in a system with a $U(1)$ gauge field where a charge $3$ object condenses and partially breaks the gauge symmetry.  Importantly, unlike free photons, this discrete gauge theory has no low lying propagating modes.  In addition, the bulk supports string-like excitations that carry flux of the $Z_3$ gauge field.  These are like vortex lines in the charge three superconductor analogy.  All of the complicated many-body physics is subsumed into the assumption of fractionalization $c \rightarrow f_1 f_2 f_3$.  The resulting fractional charges can be weakly coupled.  Thus the fractional charges may fill a band structure, and for a fractional topological insulator they fill a $Z_2$ non-trivial topological band.  We emphasize that it is the partons that may be approximated as filling a topological band, not the electrons.  The low energy effective theory contains a $\theta$ term which written in terms of the electron charge has $\theta = \pi/9$.  Nevertheless, the state is time reversal invariant, as described in Refs. \onlinecite{fti1,fti2}, because of the presence of topological ground states depending on the topology of space\cite{toporder2}.  We have a single ground state in infinite space and $3^3 = 27$ ground states with periodic boundary conditions.

If instead of considering periodic boundary conditions we consider a sample with a surface then the system has gapless surface states.  In the minimal case, these surface states consist of a single Dirac cone of charge $e^* = e/3$ quasiparticles.  Again, we emphasize that this is not a Dirac cone for electrons.  However, if the chemical potential is tuned to the Dirac point then it will still have a specific heat going like $T^2$ and will generally share many thermodynamic and transport properties with more familiar electronic surface states.  In the presence of time reversal breaking surface perturbations these surface states may be gapped out producing a Hall response.  In this case the surface Hall conductivity is
\begin{equation}
\sigma_{xy} = \frac{1}{2}\frac{(e^*)^2}{h} = \frac{1}{18}\frac{e^2}{h}
\end{equation}
which is simply the usual Dirac fermion result for charge $e^*$ fermions.  Note that this result is consistent with a bulk $\theta$ term with $\theta = \pi/9$.  Below we will describe the signatures of these surfaces in photoemission, but for the moment let us consider the effect of various perturbations to this basic story.

\section{Doping and proximate superconductivity}

Consider first the case of proximate superconductivity.  A nearby s-wave superconductor induces a term of the form $\Delta c_{r \uparrow} c_{r \downarrow}$ in the low energy effective theory of the surface.  In the case of topological insulators this term strongly modifies the low energy physics.  This may seen from the fact that near the Dirac cone, where the effective action is
\begin{equation}
\mathcal{L} = \bar{\psi} i \gamma^\mu \partial_\mu \psi
\end{equation} the electron operator $\psi \sim c$ is a scaling field with dimension one.  A mass term of the form $\bar{\psi} \psi$ as well as the superconducting term $\psi \psi$ are both of dimension two and hence relevant at the Dirac fixed point.  However, for the fractional topological insulator we are considering, the operator $c c$ is highly composite.  In terms of the weakly correlated quasiparticles we have $c c \sim f f f f f f$ and hence $c c$ is highly irrelevant at the Dirac fixed point.  It remains irrelevant when the chemical potential moves away from the Dirac point and the partons form a Fermi surface.  The term $f f$ which could gap the quasiparticles carries $Z_3$ charge and cannot be induced by the $Z_3$ gauge neutral superconducting perturbation.  Thus we have the interesting conclusion that the surface states of a fractional topological insulator are not gapped by proximity to a sufficiently weak superconductor.  Of course, it is always possible that the surface states spontaneously break their gauge symmetry in which case normal superconductivity is induced.

Next we consider the effect of doping.  Our motivation is two fold: we regard doping as particularly natural experimental knob and it appears to be an unavoidable reality in some materials.  Because the low energy quasiparticles carry conventional electric charge in addition to gauge charge, we still know how an electric potential couples to the low energy quasiparticles.  For a finite range of chemical potential, the bulk state is stable since it may be described as a band insulator of fractionally charged fermions.  On the other hand, the surface states fill in the bulk gap and may fill continuously as we vary the chemical potential.  Unless we somehow finetune the chemical potential to the Dirac point, the surface states can form a 2d Fermi liquid while the bulk remains insulating.  Surface transport provides a host of interesting signatures.  For example, if we can neglect other contributions to the thermal conductivity then the Wiedemann-Franz law should be valid, but the effective Lorenz number will be
\begin{equation}
L = \frac{\pi^2}{3} \left(\frac{k_B}{e^*}\right)^2 = 9 L_0
\end{equation}
due to the fractional charge.  We assume the surface states will form a Fermi liquid since the surface quasiparticles need not be strongly correlated.  Of course, this state may be unstable to superconductivity at low temperatures, but we ignore that possibility in the present work.

As we know from real materials, sometimes the materials are self doped in some way so that instead of directly controlling the chemical potential we simply add or remove electrons.  In this case, if we add or remove electrons in the bulk, we may form a bulk Fermi surface in addition to the boundary Fermi surface.  Like the boundary Fermi surface, however, this state can be relatively simple, essentially a 3d Fermi liquid of charge $e^*$ quasiparticles.  It will also have the thermodynamics and transport of a Fermi liquid with exotic single particle properties.

\section{Photoemission, tunneling, and quantum oscillations} Let us now turn to a discussion of the photoemission signatures of such a phase.  Photoemission has played a crucial role in the discovery of ordinary topological insulators, in particular, the surface Dirac has proven itself a very robust signature of these phases.  The most immediate consequence of fractionalization is that photoemission will no longer see a sharp quasiparticle peak.  This is not because the charge $e^*$ quasiparticles are not sharp, but rather because the electron which is ejected in photoemission is a highly composite object.  Heuristically, the probability to eject an electron from the surface is drastically reduced because we must gather together three weakly correlated partons to form an electron.

These considerations may be made precise by considering the electron spectral function as measured in photoemission.  The fractionalization ansatz $c = f_1 f_2 f_3$ together with the assumption that the quasiparticles $f_i$ interact weakly leads to an electron spectral function of the form
\begin{equation}
A(\omega,k) \sim \int d^2 q_1 d^2 q_2 d^2 q_3 \delta^2 \left(\vec{k} - \sum_i \vec{q}_i\right) \delta \left(\omega - \sum_i \epsilon_i\right)
\end{equation}
where we have specialized to the surface and the dispersion is $\epsilon_i \sim v |q_i|$.  Formally this follows from the result that the electron spectral function is a convolution of the parton spectral functions up to non-universal terms in the absence of singular vertex corrections.  Notice how the three body phase space integrals wash out the delta function in energy.  Hence there is no quasiparticle peak in the electron spectral function.  However, there are power law singularities.  We find $A(\omega,k) \sim (\omega - v|k|)^3 \theta(\omega - v|k|)$.  Distinguishing this power law edge from the incoherent background in photoemission data is a challenge for experimentalists.  Of course, the mere absence of a spectral peak in the presence of some other measurements may already be an indirect sign of the fractional topological insulator.

Even if the chemical does not sit at the Dirac point, a similar situation obtains in photoemission.  The energies entering Eq. 5 are modified to $\epsilon_i = v |q_i| - \mu$ with $\mu = v k_F$ assuming the Dirac cone remains a good description of the band structure in the relevant range of doping.  We must also require that all energies be of the same sign when carrying out the integral in Eq. 5 (this follows from a detailed derivation but is intuitively plausible).  The quasiparticle peak will be washed out and replaced with a smooth background even less singular than the Dirac cone case.  Thus one would have a very interesting situation where the surface could transport charge and heat like a Fermi liquid and yet have completely different single particle properties.

We now turn to tunneling and quantum oscillations measurements.  Using the same formalism as above, we may compute the tunneling density of states
\begin{equation}
\mbox{DOS}(\omega) \sim \int d^2 k A(\omega,k) \sim \omega^5.
\end{equation}
Thus tunnel current measurements will produce a highly non-linear $dI/dV$ curve, at least for clean samples.  Away from the Dirac point the tunneling density of states is again modified to $\mbox{DOS}(\omega) \sim \omega^2$.

We can also study the effects of charge impurities on the electron density.  As we have repeatedly emphasize, the surface states are not weakly correlated in terms of electrons but they are weakly correlated in terms of partons.  Hence the surface electrons are not in a Fermi liquid state at finite density, nevertheless, charged impurities will induce Friedel-like oscillations.  This is because the partons are in a Fermi liquid state at finite density and the parton density operator has the right quantum numbers to couple to probes of the electron density.  Because the partons are charged they will form an oscillating parton density pattern in response to a charged impurity.  The parton density pattern will then, via linear response, effect a similar pattern in any quantity that can couple to parton density.  The resulting oscillations will occur at the partonic $2 k_F$ (and harmonics), but the amplitude may be reduced relative to the usual Fermi liquid result.  We cannot reliably compute the amplitude without further information about the state.  We should point out that in principle all higher harmonics are included, and unless some harmonic of the relevant wavevector is commensurate with the reciprocal lattice, weaker and weaker singularities associated with higher harmonics will, after folding back to the first Brillouin zone, fill up the zone.

Now Friedel oscillations may be relatively hard to observe, but as we now show, combined with quantum oscillations, they offer a sharp signature of fractional charge.  Because the gapless partons form a Fermi liquid state when the surface is at generic chemical potential and because they couple to the magnetic field in the usual way through their charge, they will display quantum oscillations in magnetization, resistance, etc.  The period of these oscillations will be
\begin{equation}
\Delta\left(\frac{1}{B} \right) = \frac{2 \pi e^*}{\hbar c A_F}
\end{equation}
where $A_F$ is the area of the Fermi surface.  Crucially it is $e^*$ which appears in this formula, and so we see that if we have an independent measure of the Fermi surface area, say through Friedel oscillations, then we may calculate the ``conventional period" $\frac{ 2\pi e}{\hbar c A_F}$ and take the ratio to obtain a measure of $e^*/e$ thus extracting the fractional charge.  This prediction is quite unambiguous, however, we note that it may also be possible to obtain the fractional charge from the envelope of the quantum oscillations.  Since the partons sit in a Fermi liquid state, a detailed analysis of the temperature dependence of the oscillation envelope might also provide the quasiparticle charge via the cyclotron frequency.

\section{Discussion} We have discussed a wide array of sharp experimental probes of fractional topological insulators.  Typically these probes respond to the presence of highly non-trivial surface states which when combined with the physics of electron fractionalization lead to striking experimental signatures.  Precisely because the surface states respond to so many probes, fractional topological insulators will be amongst the easiest fractionalized phases to discover experimentally.  Indeed, all known experimental examples of fractionalized phases, from the fractional quantum Hall fluids to the 2d organics, appear to have gapless modes either in the bulk or at a boundary.

We briefly review the experimental signatures discussed in this work.  There will be magneto-electric effects similar to ordinary topological insulators although different in magnitude.  Photoemission will not show a sharp quasiparticle peak, instead there will be power law edges in addition to an incoherent background.  Tunneling experiments will observe a highly non-linear tunneling density of states.  At generic values of the chemical potential Friedel oscillations and quantum oscillations should be visible with the fractional charge $e^*$ extractable.  If the bulk is doped or we can measure surface transport, then we expect a Wiedemann-Franz type law to hold but with a highly renormalized Lorenz number.  We also predicted that unlike ordinary topological insulators, the surface states in a fractional topological insulator are perturbatively robust to proximate superconductivity.  Furthermore, the fractionalization in the bulk is robust to doping, so that the doped bulk is not smoothly connected to a Fermi liquid of electrons (unlike in conventional 3d topological insulators).

We can consider a variety of detection protocols.  A fractional magneto-electric response in a $\mathcal{T}$ invariant material is the most direct experimental probe.  We expect fractional topological insulators to give similar responses to conventional topological insulators (albeit with different $\theta$) in measurements like the Kerr rotation.  A magneto-electric response combined with the absence of a sharp quasiparticle peak in photoemission would be reasonably good circumstantial evidence for a fractional topological insulator state.  It may be hard to see the detailed photoemission structure we have described, but the robustness of the surface states to superconductivity could help.  It may be possible to significantly reduce the background in ARPES by applying a proximate s-wave superconductor to the surface, perhaps a thin film, which if sufficiently weak would lift conventional electronic states while leaving the fractionalized surface states gapless. A surface phase transition as a function strength of proximate superconductivity and a bulk phase transition as a function of doping are also sharp predictions of the theory.

We have not directly considered the effects of long range Coulomb interaction.  If there is a metallic density of quasiparticles, then the Coulomb interaction may be largely screened.  On the other hand, as is well known from graphene, the role of the Coulomb interaction at a Dirac point is quite complex.  Fortunately, because the state of the fractionally charged quasiparticles is similar to that of an ordinary topological insulator, many results will carry over rather directly.  Along these same lines, we expect that because quantum oscillations and other probes have already been successfully applied to ordinary topological insulators, there is no reason in principle why the long range Coulomb interaction will obstruct these observations.

We would also like to mention that although the surface states are not localized by disorder, the surface density of states can nevertheless be modified.  Because the low energy quasiparticles are charged they coupled to charge disorder (and phonons) essentially like ordinary electrons.  We must assume a clean sample to trust the details of our density of states calculations, and disorder can also wash out quantum oscillations.  Furthermore, polycrystalline samples may have bulk conductivity even in the absence of doping, so we should not necessarily expect truly insulating behavior from the bulk except for single crystal samples of known chemical potential.

Finally, it is possible to describe a simple continuous quantum phase transition between a fractional topological insulator and a fractionalized phase without any edge state.  Furthermore, if we break time reversal then no phase transition is needed at all.  In the presence of time reversal, this transition is completely analogous to the phase transition between a trivial and a non-trivial $Z_2$ insulator except that it is the fractionally charged quasiparticles whose band structure is changing.  The electron remains fractionalized on both sides of the phase transition and at the critical point.  Thus a detection of a fractional topological insulator would also be very interesting in that there would be proximate fractionalized phases that are fully gapped on the surface and in the bulk.  There is not yet an experimental example of such a phase of matter.

As for candidate materials, one promising family may be the iridates \cite{elec_struc_iridates,ti_iridates1} where a closely related fractionalized phase was recently proposed \cite{mott_spinorbit_pesin}.  Speaking crudely, these materials combine the raw ingredients, spin-orbit coupling and strong correlations, needed to realize a fractional topological insulator.  Note also that it is not obvious what sort of electronic band structure leads to the most robust fractional topological insulator state.  Conventional wisdom from the fractional quantum Hall effect tells us to look for nearly flat bands where the electron kinetic energy is quenched, but more detailed numerical studies suggest a more subtle picture: electron band flatness does not directly correlate with a robust many-body state (as measured by the bulk gap)\cite{fci_zoo}.  Of course, we must give interactions a chance to do their work, but we may not need to insist on extremely flat electron bands.

Where do we stand now in the search for these fascinating states?  We have provided a low energy theory that can be reliably analyzed leading to a plethora of sharp experimental predictions.  We can provide trial wavefunctions for the quantum state and we can suggest, via a strong coupling expansion \cite{strongcoupling}, terms in the electronic Hamiltonian that favor forming a fractional topological insulator.  We can also generate reliable experimental predictions for other candidate phases, including those with gapless gauge fields \cite{mott_spinorbit_pesin,gaugefluc_tmi}, that may also be realized in strongly interacting spin-orbit materials.  The one thing we cannot yet provide is a really sharp prediction for a candidate material, but work is in progress to provide a more refined picture.

\textit{Acknowledgements} I thank J. McGreevy, T. Senthil, M. Barkeshli, and D. Hsieh for stimulating discussions.  This work was supported by a Simons Fellowship through Harvard University.

\bibliography{fti_exp}
\end{document}